\documentclass[aps,prl,twocolumn,superscriptaddress,preprintnumbers]{revtex4}
\usepackage{graphicx}
\usepackage{amssymb,amsmath}
\usepackage{mathrsfs}
\usepackage{slashed}
\usepackage{indentfirst}
\usepackage{float}
\usepackage{xcolor}
\usepackage{dsfont}
\usepackage{ulem}

\usepackage[colorlinks,
            linkcolor=black,
            anchorcolor=black,
            citecolor=black
            ]{hyperref}

\begin{document}


\title{Models with higher weak-isospin Higgs multiplets}

\author{Cheng-Wei Chiang}
\email[e-mail: ]{chengwei@phys.ntu.edu.tw}
\affiliation{Department of Physics, National Taiwan University, 
Taipei, Taiwan 10617, R.O.C.}
\affiliation{Institute of Physics, Academia Sinica, Taipei, Taiwan 11529, R.O.C.}
\affiliation{Kavli IPMU, University of Tokyo, Kashiwa, 277-8583, Japan}

\author{Kei Yagyu}
\email[e-mail: ]{yagyu@st.seikei.ac.jp}
\affiliation{Seikei University, Musashino, Tokyo 180-8633, Japan}


\begin{abstract}
In order for scale factors $\kappa_{V}^{}~(V=W,Z)$ of the 125-GeV Higgs boson couplings to have the possibilities of being greater than unity and $\kappa_{W}^{} \neq \kappa_{Z}^{}$
while keeping the electroweak $\rho$ parameter unity at tree level, 
the Higgs sector must be extended with at least two exotic $SU(2)_L$ multiplets in addition to the doublet Higgs field in the Standard Model.  
By the requirements of perturbative unitarity, no Landau pole in gauge couplings, and no accidental global $U(1)$ symmetry, 
we exhaust all the possible combinations of two exotic Higgs fields and derive general formulas for $\kappa_{V}^{}$.  
We find that the current central values $\kappa_W^{}=1.12$ and $\kappa_Z = 0.99$ reported by CMS can be accommodated in the model with a complex and a real Higgs triplets as the simplest example. 
\end{abstract}


\maketitle


{\it Introduction ---}
In the pursue of models with an extended Higgs sector, an important empirical constraint is the electroweak $\rho$ parameter.  
Its experimental value has been known to be quite close to unity, {\it i.e.}, $1.00039 \pm 0.00019$~\cite{pdg}.  
This fact suggests that the extended Higgs sector should naturally be constructed so as to have $\rho = 1$ at tree level. 

In addition to the $\rho$ parameter, 
measurements of 125-GeV Higgs boson $(h)$ couplings to a pair of Standard Model (SM) particles $X$
are important to further narrow down the structure of the Higgs sector.  
To discuss the compatibility, 
the scale factor $\kappa_X^{}$ defined as the $hXX$ coupling normalized to its SM prediction is often introduced. 
If deviations in the $h$ couplings, {\it i.e.}, $\kappa_X^{} \neq 1$ are established in future experiments, 
they will be indirect evidence of physics beyond the SM, particularly the existence of an extended Higgs sector.

Currently, measured values of $\kappa_W^{}$ and $\kappa_Z^{}$ at the CERN Large Hadron Collider (LHC) are summarized in Table~\ref{tab:kappas}.
It is interesting to note that the central values are ``non-standard,'' suggesting that not only could they be greater than unity, there is also a possibility that they differ from each other by $\sim 10\%$ according to the CMS result.  Even though these possibilities are far from conclusive due to the large uncertainties on both quantities at the moment, it is anticipated that the errors will reduce to $2 - 4\%$ at the high-luminosity LHC~\cite{Khachatryan:2016vau} or even down to the sub-percent level at the International Linear Collider~\cite{Dawson:2013bba}.  
Moreover, recently a new method has been proposed to determine $\kappa_W^{}/\kappa_Z^{}$, including its sign, at lepton colliders~\cite{Chiang:2018fqf}.
Therefore, differences between $\kappa_W$ and $\kappa_Z$ and/or from unity at ${\cal O}(10\%)$ level can become significant then. 
If such non-standard $\kappa_{W}^{}$ and $\kappa_Z^{}$ are established in the future experiments, a natural question is what kind of models features such properties.

\begin{table}[b]
\begin{tabular}{cccc}
\hline\hline
Parameter & ATLAS & CMS & Average
\\
\hline
$\kappa_W^{}$ & ~~$1.07 \pm 0.10$~~ & ~~$1.12^{+0.13}_{-0.19}$~~ & ~~$1.08 \pm 0.08$~~
\\
$\kappa_Z^{}$ & ~~$1.07 \pm 0.10$~~ & ~~$0.99 \pm 0.11$~~ & ~~$1.03 \pm 0.07$~~
\\
\hline\hline
\end{tabular}
\caption{Best-fit values and $\pm1\sigma$ uncertainties of $\kappa_{W}^{}$ and $\kappa_{Z}^{}$
reported by ATLAS and CMS Collaborations~\cite{ATLAS:2018doi,CMS:2018lkl} using  
13-TeV collision data and with $79.8$~fb$^{-1}$ and $35.9$~fb$^{-1}$ integrated luminosities, respectively. 
Numbers shown here are based on the assumption of no beyond the SM decay channels for $h$.  The last column gives the weighted averages of the two quantities, where the errors have been symmetrized.
\label{tab:kappas}}
\end{table}

It has been known that $\kappa_V^{} > 1$ ($V=W,Z$) cannot be realized in a Higgs sector 
constructed with only isospin doublets and/or singlets, but is possible in models having triplets or higher multiplets~\cite{KKY,Chiang:2013rua}. 
However, if we impose the custodial symmetry in the Higgs sector in order to keep $\rho = 1$ at tree level, such as in the Georgi-Machacek (GM) model~\cite{GM}, $\kappa_W^{} = \kappa_Z^{}$ is also predicted at tree level. 
Interestingly, requiring $\rho = 1$ at tree level does not necessarily call for the custodial symmetry.  
Moreover, a sizable difference between $\kappa_W^{}$ and $\kappa_Z^{}$ can be accommodated in the case without the custodial symmetry.

{\it Scenarios ---}
Inspired by the above observations, we explore plausible extended Higgs sectors that have $\rho = 1$ at tree level, allow $\kappa_{V}^{}$ to be greater than 1 as well as predict sufficiently different values for them.  

Let us consider a renormalizable extended Higgs sector composed of a Higgs doublet $\Phi$ as in the SM and $N$ extra Higgs multiplets $X_a$ ($a = 1, \cdots, N$)
whose $SU(2)_L$ and $U(1)_Y$ quantum numbers are $(T_a,Y_a)$. 
The electric charge of a particular component field in $X_a$ is given by $Q_a = T_a^3 + Y_a$ with $T_a^3$ denoting the third component of weak isospin. 
In order for each of $X_a$ to participate in electroweak symmetry breaking, 
it must have a neutral component and, therefore, $Y_a$ must be an integer (half integer) for integer (half integer) $T_a$.  
The vacuum expectation value (VEV) of the neutral component of $X_a$ is denoted as $v_a/\sqrt2$ ($v_a$) if it is a complex (real) scalar.  
For simplicity, we do not consider either explicit or spontaneous CP violation in the Higgs potential.  

It is known that introduction of a Higgs multiplet with too large $T_a$ breaks perturbative unitarity of tree-level scattering amplitudes, 
{\it e.g.}, the Scalar-Scalar $\to$ Gauge-Gauge type of processes~\cite{Hally:2012pu}
due to the enhancement in Scalar-Scalar-Gauge couplings. 
According to Ref.~\cite{Hally:2012pu}, the maximum size of $T_a$ is given by $7/2~(4)$ for a complex (real) scalar in the $N = 1$ case.
We thus impose the same upper limits on $T_a$, although more severe conditions would be obtained for larger $N$ because of additional contributions to the scattering processes.

In such a model, the electroweak $\rho$ parameter, defined by $\rho
\equiv m_W^2 / (m_Z^2 \cos^2\theta_W)$ with $m_W^{}~(m_Z^{})$ and $\theta_W$ denoting respectively the $W$ $(Z)$ boson mass and the weak mixing angle, at tree level is given by
\begin{align}
\rho_{\text{tree}}
=
\frac{v_\Phi^2 + 2 \sum_{b = 1}^N v_b^2 [T_b (T_b + 1) - Y_b^2]}
{v_\Phi^2 + 4 \sum_{a = 1}^N v_a^2 Y_a^2}
~,
\end{align}
where $v_\Phi = \sqrt{2}\langle \Phi^0 \rangle$.  The condition $\rho_{\rm tree} = 1$ gives
\begin{align}
\sum_{a = 1}^N v_a^2 [T_a (T_a + 1) - 3Y_a^2] = 0~.
\label{eq:rho=1}
\end{align}
In the case of $N = 1$, we have the well-known solutions: 
\begin{align}
(T_1,Y_1) = (0,0),~~(1/2,1/2),~~(3,2). \label{comb}
\end{align}
As already mentioned above, the first two solutions always have $\kappa_{V}^{} \le 1$, while the last one allows the possibility $\kappa_{V}^{} > 1$~\cite{Hisano,KKY}.  
However, all of these solutions predict $\kappa_W = \kappa_Z$ at tree level, 
which is expected to be violated at ${\cal O}(1\%)$ level or smaller when radiative corrections are included (see, for example, Ref.~\cite{Chiang:2017vvo}).

We therefore consider the next simplest case of $N = 2$.  Without loss of generality, we assume that $T_1 \ge T_2$.  
Moreover, neither multiplets have the quantum numbers shown in Eq.~(\ref{comb}); otherwise,
it reduces to a model with $\kappa_{W} = \kappa_{Z}$ at tree level.  
With Eq.~\eqref{eq:rho=1}, the two VEVs satisfy the following relation: 
\begin{align}
r \equiv \frac{v_2^2}{v_1^2} = - \frac{T_1 (T_1 + 1) - 3Y_1^2}{T_2 (T_2 + 1) - 3Y_2^2} 
~.
\label{eq:r}
\end{align}
Besides, we have a sum rule about the VEVs:
\begin{align}
v^2 = v_\Phi^2 + \xi^2 v_1^2 
~~~~\mbox{with}~~
\xi^2
\equiv 
4 \left( Y_1^2 + r Y_2^2 \right)~,
\label{eq:xi}
\end{align}
where $v \simeq 246$~GeV.
For later convenience, we define a mixing angle $\beta$ through
\begin{align}
\tan\beta = \frac{v_\Phi}{\xi v_1} \label{tanb}
\end{align}
in a way consistent with that in two-Higgs doublet models~\cite{HHG}.  More explicitly, 
$\beta$ is the mixing angle to separate the Nambu-Goldstone (NG) bosons to be absorbed into the weak gauge bosons from the physical Higgs states such as CP-odd and singly-charged Higgs bosons.

Even if models with $N = 2$ satisfy the condition given in Eq.~(\ref{eq:rho=1}), some of them should be excluded because of the existence of 
accidental global $U(1)$ symmetries associated with phase rotations of $X_1$ and $X_2$ 
as they would give rise to at least one phenomenologically unacceptable massless NG boson after the electroweak symmetry breaking. 
To avoid such NG bosons, we require that there be no accidental global $U(1)$ symmetry in the Higgs potential.  The following are general situations where such a $U(1)$ symmetry can be broken explicitly by renormalizable terms.
First, introducing a multiplet of $(T_a,Y_a)=(1,0),~(1,1),~(3/2,1/2)$ and $(3/2,3/2)$ is safe because they can couple with an appropriate number of $\Phi$ fields and/or their conjugates.  Also, having a real multiplet $(Y = 0)$ is all right because its bilinear and higher power terms can be constructed.  Finally, for a multiplet other than the above-mentioned ones, one needs to check if at least one renormalizable term involving this multiplet and the other scalars can be constructed.

We then find all possible scenarios, as listed in Table~\ref{tab:summary}.  
Some of them would result in a Landau pole in the $SU(2)_L$ gauge coupling below the Planck scale.  The scale at which the Landau pole appears, denoted by $\Lambda_{\rm LP}$, is 
calculated by using one-loop renormalization group equations and given in the table. 
We note that the Landau pole sometimes also appears in the $U(1)_Y$ gauge coupling. But its scale is always higher than that of the $SU(2)_L$ gauge coupling. 
In addition, we also show the upper limit on $v_1$, denoted by $v_1^{\rm max}$, 
by requiring that the top Yukawa coupling $y_t$ remains perturbative, {\it i.e.}, $y_t \le \sqrt{4\pi}$ at the electroweak scale.

\begin{table}[t]
\begin{ruledtabular}
\begin{tabular}{cccccc}
$(T_1,Y_1)$ & $(T_2,Y_2)$ & $r$ & $\xi^2 $ & $v_1^{\rm max}$ & $\Lambda_{\text{LP}}$
\\
\hline
(1,1) & (1,0)                   
& $1/2$ & $4$  & $118$ & --
\\
(3/2,1/2) & (1,1)                
& $3$ & $13$ & $65$ & --
\\
(3/2,3/2) & (1,0)
& $3/2$ & $9$ & $79$ & --
\\
(3/2,3/2) & (3/2,1/2)            
& $1$ & $10$ & $75$ & --
\\
(2,0) & (1,1)                    
& $6$ & $24$ & $48$ & --
\\
(2,0) &  (3/2,3/2)               
& $2$ & $18$ & $56$ & --
\\
(2,1) & (1,1)                    
& $3$ & $16$ & $59$ & --
\\
(2,1) & (3/2,3/2)                
& $1$ & $13$ & $65$ & --
\\
(2,2) & (2,1)                
& $2$ & $24$ & $48$ & --
\\
$(5/2,1/2)$ & $(1,1)$            
& $8$ & $33$ & $41$ & --
\\
$(5/2,1/2)$ & $(3/2,3/2)$        
& $8/3$ & $25$ & $47$ & --
\\
$(5/2,3/2)$ & $(1,1)$            
& $2$ & $17$ & $57$ & --
\\
$(5/2,3/2)$ & $(3/2,3/2)$        
& $2/3$ & $15$ & $61$ & --
\\
$(5/2,5/2)$ & $(5/2,3/2)$        
& $5$ & $70$ & $28$ & $2.9\times 10^{11}$
\\
(3,0) & (1,1)                    
& $12$ & $48$ & 34 & --
\\
(3,0) & (3/2,3/2)                
& $4$ & $36$ & 39 & --
\\
(3,1) & (1,1)                
& $9$ & $40$ & 37 & $5.9\times 10^{13}$
\\
(3,1) & (3/2,3/2)                
& $3$ & $31$ & $42$ & $1.9\times 10^{12}$
\\
(3,1) & (2,2)                
& $3/2$ & $28$ & $45$ & $2.9\times 10^{10}$
\\
(7/2,1/2) & (1,1)                
& $15$ & $61$ & $30$ & $9.9\times 10^8$
\\
(7/2,1/2) & (3/2,3/2)          
& $5$ & $46$ & 35 & $2.7\times 10^{8}$
\\
(7/2,3/2) & (3/2,3/2)          
& $3$ & $36$ & 39 & $2.7\times 10^{8}$
\\
(7/2,3/2) & (5/2,5/2)                
& $9/10$ & $63/2$ & $42$ & $6.7\times 10^6$
\\
(7/2,5/2) & (5/2,3/2)          
& $3/2$ & $77/2$ & 38 & $6.7\times 10^{6}$ 
\\
(7/2,5/2) & (7/2,3/2)          
& $1/3$ & $28$ & 45 & $1.7 \times 10^5$ 
\\
(4,0) & (1,1)
& 20 & 80 & 26 & $5.3\times 10^{12}$
\\
(4,0) & (3/2,3/2)
& 20/3 & 60 & 30 & $2.9\times 10^{11}$
\end{tabular} 
\end{ruledtabular}
\caption{Viable scenarios with $N = 2$.  The first two columns give the $SU(2)_L$ and $U(1)_Y$ quantum numbers of $X_1$ and $X_2$, respectively.  
The third and fourth columns give values of $r$ and $\xi^2$ defined in Eqs.~\eqref{eq:r} and \eqref{eq:xi}, respectively.  
The fifth column gives the maximally allowed value of $v_1$ by demanding perturbativity on the top Yukawa coupling.
The last column gives the scale at which a Landau pole in the running of the $SU(2)_L$ coupling appears below the Planck scale.  
The unit of the last two columns is GeV.
}
\label{tab:summary}
\end{table}

In order to obtain the expressions for the Higgs boson couplings, let us denote the three CP-even scalars associated with $(\Phi,X_1,X_2)$ as $(h_\Phi,h_{X_1},h_{X_2})$ 
and the three physical states as $(h,H_1,H_2)$.  The two sets of fields are related by
\begin{align}
\begin{pmatrix}
h_\Phi \\ h_{X_1} \\ h_{X_2}
\end{pmatrix}
=
R
\begin{pmatrix}
h \\ H_1 \\ H_2
\end{pmatrix}
~,
\end{align}
where $R$ is a $3 \times 3$ orthogonal rotation matrix. 
In general, $R$ involves three independent mixing angles.  
Note that if we impose the custodial symmetry in the Higgs potential, the structure of the matrix $R$ is more constrained. 
For example, in the GM model, two of the three mixing angles are determined by the symmetry and it results in $\kappa_W^{} = \kappa_Z^{}$ at tree level (see, {\it e.g.}, Ref.~\cite{Chiang:2017vvo}). 

In this Letter, we do not assume the custodial symmetry in the potential and, hence, $R$ takes the most general form with three arbitrary mixing angles. 
We then obtain the general expressions for $\kappa_W^{}$ and $\kappa_Z^{}$ as 
\begin{align}
\begin{split}
\kappa_W^{}
&= 
s_\beta R_{11} + c_\beta \frac{2 [T_1 (T_1 + 1) - Y_1^2] R'}{\xi} c_\theta
\\
& \qquad\qquad
+ c_\beta \sqrt{r} \frac{2 [T_2 (T_2 + 1) - Y_2^2] R'}{\xi} s_\theta
~,
\\
\kappa_Z^{}
&=
s_\beta R_{11} + c_\beta \frac{4 Y_1^2 R'}{\xi} c_\theta + c_\beta \sqrt{r} \frac{4 Y_2^2 R'}{\xi} s_\theta
~,
\end{split} \label{kappa_v}
\end{align}
where $R' \equiv \sqrt{1 - (R_{11})^2}$, $\theta \in [0,2\pi)$, 
and the shorthand notation $c_\phi = \cos\phi$ and $s_\phi = \sin\phi$ for $\phi = \beta,\theta$ has been introduced.  
We note that $\kappa_W^{} = \kappa_Z^{}$ is obtained when
\begin{align}
\tan\theta = - \sqrt{r}~.
\end{align}
On the other hand, only the doublet Higgs field $\Phi$ couples to the SM fermions, so that the scale factor for Yukawa couplings ($\kappa_F^{}$) is universally given by
\begin{align}
\kappa_F = \frac{R_{11}}{s_\beta}~. \label{kappa_f}
\end{align}

{\it Example ---}
Here we work out the simplest scenario of the $N = 2$ models; {\it i.e.}, the model with $X_1 \sim (1,1)$ and $X_2 \sim (1,0)$.  
This corresponds to the generalized version of the GM model without the custodial symmetry.  

\begin{figure}[t!]
\centering{
\includegraphics[width=0.9\columnwidth]{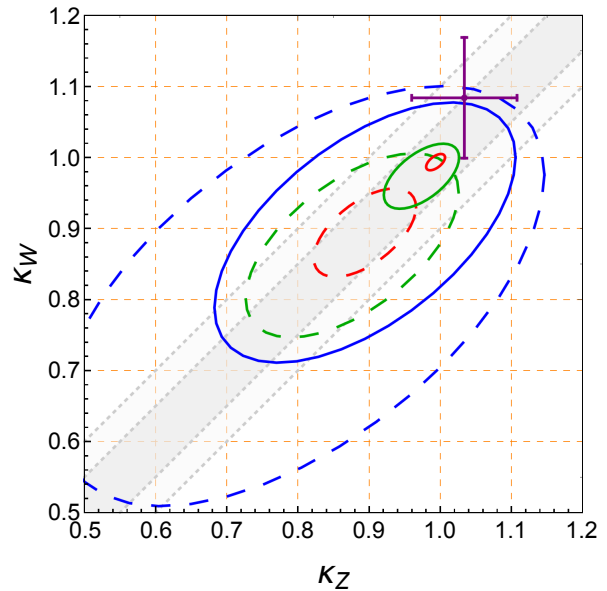}
}
\caption{Correlation of $\kappa_Z^{}$ and $\kappa_W^{}$ in the model with $X_1\sim (1,1)$ and $X_1\sim (1,0)$.  
The dashed and solid curves respectively show the cases for $\kappa_F = 0.9$ and $1.0$ with $v_1 = 10$ (red), 20 (green) and 40~GeV (blue). 
For each ellipse, $\theta$ is scanned from 0 to $2\pi$. 
The dark (light) gray band indicates $|\kappa_Z - \kappa_W| \le 5\%$ ($10\%$).  The purple cross marks the current weighted data from Table~\ref{tab:kappas}.}
\label{fig:kZkW}
\end{figure}

In Fig.~\ref{fig:kZkW}, we show the correlation between $\kappa_Z^{}$ and $\kappa_W^{}$ for fixed values of $\kappa_F^{}$: 0.9 (dashed curves) and $1$ (solid curves).  
The red, green and blue curves correspond to the representative cases of $v_1 = 10$, 20 and 40~GeV, respectively.  
From Eq.~(\ref{kappa_f}), fixing $\kappa_F^{}$ and $v_1$, or equivalently $\tan\beta$ via Eq.~(\ref{tanb}), determines $R_{11}$. 
Then, $\kappa_W^{}$ and $\kappa_Z^{}$ are determined as functions of the angle $\theta$. By scanning $\theta$ from 0 to $2\pi$, 
we obtain the ellipses shown in this figure.  
It is seen that for the case with larger $1-\kappa_F$ and/or larger $v_1$, the size of the ellipse becomes larger. 
This is because contributions from the second and third terms in Eq.~(\ref{kappa_v}) become more significant 
as $R' \cos\beta$ gets larger. 
We find that an ${\cal O}(10\%)$ difference between $\kappa_W^{}$ and $\kappa_Z^{}$ 
can be achieved in each value of $\kappa_F^{}$ with some appropriate choice of $v_1$ whose corresponding ellipse can intersect with the boundaries of the light gray area.  
To fit exactly the central values of CMS data, one should choose $v_1 \simeq 44$ and 50~GeV for $\kappa_F = 0.9$ and 1.0, respectively. 

\begin{figure}[t]
\centering{
\includegraphics[width=0.9\columnwidth]{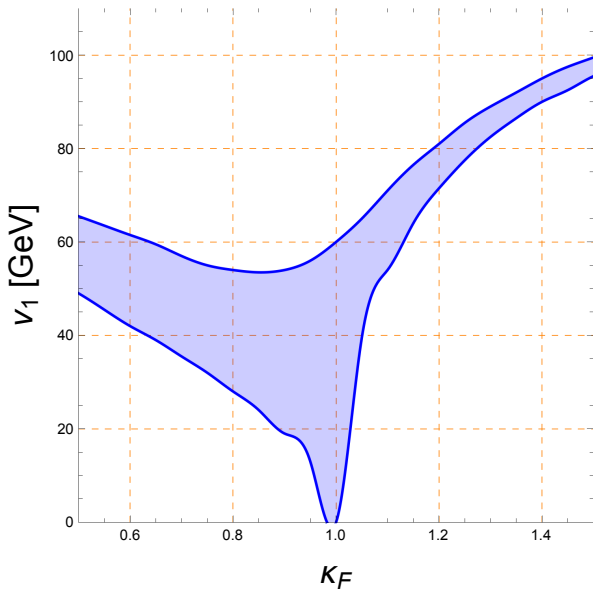}
}
\caption{Region in the $\kappa_F$-$v_1$ plane allowed by the averaged $\kappa_{W}^{}$ and $\kappa_{Z}^{}$ given in Table~\ref{tab:kappas}.}
\label{fig:paraspace}
\end{figure}

By requiring $\kappa_{V}^{}$ to fall within their respective averaged $1\sigma$ ranges given in Table~\ref{tab:kappas}, 
one can obtain allowed regions on the $\kappa_F$--$v_1$ plane as shown in Fig.~\ref{fig:paraspace}.  
For each specific value of $\kappa_F$, there are upper and lower bounds on $v_1$ such that the corresponding ellipses can render the required $\kappa_{V}^{}$.  
Such a range is expected to shrink if the errors on $\kappa_V^{}$ are reduced.  
The lower bound on $v_1$ is higher for $\kappa_F > 1$ because $\sin\beta$ has to be sufficiently small to keep $R_{11} \le 1$ in Eq.~\eqref{kappa_f}.

Finally, let us consider the constraints from the electroweak $S$ and $T$ parameters in the above model. 
First of all, the $T$ parameter is not calculable in models with Higgs multiplets other than those shown in Eq.~(\ref{comb}), 
since an additional counterterm, with respect to the SM case, appears in the electroweak sector.  
In fact, this counterterm successfully cancels the ultra-violet divergence in the $T$ parameter.  But an arbitrariness remains in its finite part. 
Thus, $T$ is a parameter that one can fit to by tuning the counterterm. 
On the other hand, the $S$ parameter is calculable as in the SM. 
The current global fit result for the $S$ and $T$ parameters is given by 
$S =0.02 \pm 0.07$ and $T =0.06 \pm 0.06$ when fixing $U = 0$ with a correlation factor of $+92\%$~\cite{pdg}, where the SM prediction is set as $S = T = 0$.   
If we take $T = 0$, then the 95\% confidence level interval for $S$ is given by $-0.11\leq S \leq 0.02$. 
We have checked that all the parameter choices made in Fig.~\ref{fig:kZkW} are allowed by the constraint of $S$ while setting $T = 0$, assuming that all the masses of extra Higgs bosons are 500~GeV. 
We have also verified that the predicted value of the $S$ parameter does not change much if the degenerate mass is varied from 100 to 1000~GeV. 

{\it Summary ---} 
We have discussed models with an extended Higgs sector satisfying $\rho = 1$ at tree level while allowing $\kappa_W^{}$ and $\kappa_Z^{}$ to have different values and the possibility of being greater than unity. 
In the renormalizable theory, such a Higgs sector is realized by introducing at least two exotic Higgs fields with their weak isospin representations equal to or higher than the triplet. 
Under the basic theoretical requirements, {\it i.e.}, tree-level perturbative unitarity, no accidental $U(1)$ symmetry in the Higgs potential and no Landau pole in the gauge couplings, 
we have found fifteen possible scenarios of the two exotic Higgs fields, 
as listed in Table~\ref{tab:summary}.  The condition $\rho_{\rm tree} = 1$ is guaranteed through an appropriate combination of the exotic Higgs vacuum expectation values.
It goes without saying that one can also consider similar scenarios with more than two exotic Higgs multiplets with $\rho_{\rm tree} = 1$ and obtain $\kappa_W^{} \neq \kappa_Z^{}$. 

As an explicit example that can accommodate an ${\cal O}(10\%)$ difference between $\kappa_W^{}$ and $\kappa_Z^{}$ as suggested 
by the CMS data, we have considered the model with a complex and a real isospin triplets without the custodial symmetry.  
We have shown that the ${\cal O}(10\%)$ difference can be explained by taking appropriate triplet vacuum expectation values, depending on the value of $\kappa_F^{}$. 
If such a difference between $\kappa_W^{}$ and $\kappa_Z^{}$ is significantly established in future collider experiments, such as the high-luminosity LHC and electron-positron colliders, 
the extended Higgs scenarios found in this Letter will serve useful as a guide for candidates of new physics beyond the SM. 

{\it Acknowledgments ---}
The authors would like to thank Shinya Kanemura for fruitful discussions concerning the current measurements of the Higgs boson couplings at the LHC. 
This research was supported in part by the Ministry of Science and Technology of Taiwan under Grant No.\ MOST 104-2628-M-002-014-MY4.

\end{document}